# Deep learning-based virtual perturbation reveals increasing heterogeneity of gene regulatory responses with aging

Yongjian He[1,2,3], Julia Hanna Haddad Guez[1], Xu-Wen Wang[4,#] ,Orr Levy[1,#]

[1]*Faculty of Engineering, Bar-Ilan University, Ramat Gan, 52900, Israel*

[2] *Department of Systems Science, Faculty of Arts and Sciences, Beijing Normal University, Zhuhai, 519087, China*

[3] *School of Systems Science, Beijing Normal University, Beijing, 100875, China*

[4]*Channing Division of Network Medicine, Department of Medicine, Brigham and Women's Hospital and Harvard Medical School, Boston, MA 02115, USA*

[#]*These authors jointly supervised this work. Correspondence should be addressed to O.L. (levyorr@biu.ac.il) and X.-W.W. (spxuw@channing.harvard.edu).*

## Abstract

A central hypothesis in aging biology proposes that the gradual accumulation of stochastic molecular damage drives transcriptional deregulation (TD), ultimately leading to functional decline. However, existing single-cell analyses largely rely on static expression measurements and provide limited resolution of how regulatory responses change at the level of individual cells. Here, we leverage DKO, a deep learning framework that predicts cell-specific virtual gene perturbation effects from unperturbed transcriptomes, capturing cell-to-cell differences in regulatory context without requiring a reconstructed gene regulatory network or matched perturbation data. Applying DKO to single-cell transcriptomic datasets from humans, mice, Drosophila, and C. elegans revealed that the transcriptional consequences of perturbing the same gene became increasingly heterogeneous across cells with age, indicating that gene regulatory responses become less consistent and more dependent on cellular context. Furthermore, we find that cells with higher somatic mutation burdens have greater knockout-response variability, linking accumulated molecular damage to regulatory inconsistency. Within the same cellular populations, higher impact variability was also associated with higher senescence scores, linking transcriptional dysregulation to senescence-associated states. Our findings identify a progressive loss of regulatory robustness at both the gene and cell levels as a conserved feature of aging, linking stochastic molecular damage to increasing heterogeneity in cellular regulatory responses.

# Introduction

Aging is characterized by the progressive accumulation of stochastic molecular damage that challenges cellular homeostasis and ultimately contributes to functional decline[1,2]. Genomic instability and epigenetic alterations are established hallmarks of aging, while somatic mutations and epimutations accumulate throughout life as a consequence of endogenous processes and environmental exposures[2,3]. Because these alterations arise stochastically, different cells accumulate distinct patterns of molecular damage[4,5]. Therefore, a central unresolved question is how heterogeneous and largely random alterations at the level of individual cells can converge on the recurrent functional deterioration that characterizes aging[6].

Gene regulatory networks (GRNs) provide a plausible mechanistic bridge between stochastic molecular damage and systematic cellular dysfunction[7]. GRNs organize interactions among genes and their regulators, coordinate transcriptional programs, stabilize cellular identity, and help maintain cellular functions in the presence of internal and external perturbations[8]. Their architecture can confer robustness through feedback, modularity, redundancy, and distributed regulation, while specific regulatory motifs determine properties such as response speed, persistence, and fluctuations [8,9]. Thus, the impact of molecular damage on regulatory components is not local, as perturbations can propagate through interconnected regulatory circuits and influence broader transcriptional programs. Progressive disruption of this regulatory architecture could weaken the capacity of cells to buffer molecular perturbations, converting initially local and stochastic damage into increasingly widespread transcriptional abnormalities[10].

This reasoning motivates the transcriptional deregulation (TD) hypothesis of aging, in which accumulated molecular damage, such as mutations and epimutations, progressively compromises transcriptional regulation and thereby contributes to age-associated functional decline[3,11]. Advances in single-cell transcriptomics have made it possible to examine this process at the level where stochastic damage occurs: individual cells. A widely studied manifestation of TD is increased cell-to-cell transcriptional heterogeneity. Increased variability in gene expression has been reported in aged mouse cardiomyocytes, immune cells following stimulation, human pancreatic cells, and other aging tissues[11,12]. These observations suggest that aging can be accompanied by a loss of transcriptional consistency among otherwise related cells.

However, conventional measures of transcriptional heterogeneity provide an incomplete description of TD. Differential expression captures changes in mean transcriptional state but cannot distinguish adaptive or compensatory responses from regulatory deterioration. Likewise, increased expression variance does not necessarily imply failure of the underlying regulatory system, because transcriptional variability can arise from intrinsic gene-expression stochasticity, changes in cellular state or composition, and technical variation. Indeed, systematic comparisons across aging single-cell datasets have shown that conventional measures of transcriptional noise do not consistently increase across tissues and cell types[13]. Thus, TD cannot be defined simply as increased variability in the expression of individual genes.

A more system-level view is provided by gene-to-gene transcriptional coordination. Rather than asking whether individual genes become more variable, this approach asks whether genes continue to exhibit coordinated transcriptional behavior across cells. Recently, we showed that global gene-to-gene coordination declines with age across diverse organisms, tissues, and cell types[14]. Importantly, reduced coordination was also associated with a higher burden of an age-related somatic mutational signature and with radiation-induced DNA damage[14]. These findings support the idea that aging-associated TD may involve a broad deterioration of transcriptional organization and provide a link between accumulated molecular damage and loss of regulatory coordination. Yet global gene-to-gene coordination leaves important aspects of TD unresolved. A population-level decline in transcriptional coordination cannot show how strongly dysregulation differs among genes or among individual cells. Because molecular damage accumulates stochastically, individual cells may gradually develop different regulatory states, leading to increasing divergence within an aging population (**Fig. 1a**). We therefore asked whether this loss of regulatory coherence could also be detected through the effects of gene perturbation. If gene regulatory networks become more heterogeneous with age, perturbing the same gene should produce increasingly different effects across cells, while cells with increased GRN heterogeneity, possibly due to greater aging-related molecular damage, should show greater divergence in their overall gene-impact profiles (**Fig. 1b**). Examining TD at these levels could therefore provide a more detailed view of how transcriptional regulation changes during aging.

Here, we leverage DKO[15], a deep learning framework that uses transcriptome-wide virtual gene perturbations to study TD at gene and single-cell resolution. Rather than explicitly reconstructing GRN topology, DKO learns the relationship between cellular gene states and steady-state transcriptional profiles. By virtually perturbing individual genes and propagating these perturbations through the learned model, DKO estimates the transcriptional impact of each gene in each cellular context. Because these predicted impacts depend on the transcriptional state of individual cells, they allow differences in regulatory behavior to be measured across genes and cells. This framework therefore provides a gene- and cell-resolved view of regulatory behavior underlying TD.

Applying DKO across aging single-cell transcriptomic datasets from humans, mice, Drosophila, and *C. elegans* revealed a increased virtual KO impact variability across single cells, suggesting a conserved age associated loss of gene regulatory network coherency across cells. The predicted transcriptional consequences of perturbing the same gene became increasingly heterogeneous across cells with age, indicating that gene regulatory influence becomes less consistent and increasingly dependent on cellular context. We then, further asked what drives this increased transcriptional perturbation-response variability and found that cells with higher somatic mutational burdens exhibited greater perturbation-response variability, further associating accumulated molecular damage with regulatory instability at the single-cell level. Similarly, cells with higher senescence levels exhibited greater gene- and cell-level perturbation-response variability, including within the same cellular population, further linking senescence to reduced regulatory robustness. These results suggest that aging-associated TD involves a progressive loss of regulatory consistency at the level of individual genes and cells. DKO therefore provides a framework for quantifying how the robustness of transcriptional regulation deteriorates during aging and for investigating how stochastic molecular damage may be translated into heterogeneous cellular dysfunction.

# Results

### DKO enables transcriptome-wide virtual perturbation profiling at single-cell resolution

Although experimental perturbation remains essential for functional validation, profiling perturbation responses across many genes and large numbers of individual young and aged cells is not yet feasible at scale. We therefore leveraged a data-driven knockout (DKO) framework to predict single-cell gene expression profiles and knockout impacts directly from unperturbed transcriptomic data, without requiring experimentally perturbed data for model training[15]. DKO adopts a top-down, data-driven strategy rather than explicitly reconstructing complete gene-to-gene regulatory networks. Given single-cell gene expression data from a defined cell type, DKO uses a deep-learning model to learn the mapping between a cell's gene assemblage and its corresponding gene expression profile (see **Methods**). This approach leverages naturally observed variation in gene-expression space to learn the implicit relationships that constrain cellular expression states, thereby providing a data-driven representation of the expression landscape of a specific cell type.

After training, we performed single-cell virtual knockout experiments by removing an individual target gene from the gene assemblage of each cell and using the trained model to predict the corresponding post-knockout expression profile (**Fig. 1c**). The knockout impact was quantified as the Bray-Curtis dissimilarity between the predicted post-knockout profile and the original pre-knockout profile. Repeating this procedure across cells and expressed genes enabled systematic, transcriptome-wide virtual knockout at single-cell resolution. DKO can be applied across species, tissues and cell types for which suitable single-cell expression data are available. It therefore provides a framework for testing the aging-associated transcriptional deregulation hypothesis at both the gene and single-cell levels.

**Gene perturbation-response variability increases with aging**

To examine the transcriptional deregulation hypothesis at the gene level, we quantified the variability of gene perturbation effects across individual cells. DKO assigns an impact value to each gene-cell pair $(g, c)$, representing the predicted transcriptomic effect of perturbing a given gene $g$ in a specific cell $c$. For each gene, we defined gene-level perturbation-response variability as the median absolute deviation of its predicted impact across cells (see **Methods**). Higher variability therefore indicates that perturbing the same gene produces more heterogeneous effects among individual cells.

We first investigated gene-impact variability relationship using GRN dynamical simulations, in which individual cells were initialized from a common baseline GRN and subsequently subjected to stochastic regulatory damage (Fig. **S1a**). Each regulatory edge was independently subjected to damage with probability p. Increasing p from 0 to 0.15 progressively increased the extent of regulatory damage, allowing the GRNs of individual cells to diverge from their shared initial state and from one another. To obtain ground-truth perturbation impacts, we individually knocked out each gene in each simulated cell by setting the target gene to zero and rerunning the GRN dynamics to steady state. The resulting expression profile was then compared with its corresponding unperturbed profile (**see Methods**). Ground-truth gene impact variability increased with regulatory damage (Fig. **S1b**), indicating that perturbing the same gene produced increasingly divergent effects across cells. Importantly, using only the corresponding unperturbed expression profiles, DKO successfully predicted the same increase in gene impact variability with increasing regulatory damage (**Fig. S1c**), closely matching the trend obtained from the ground-truth perturbation effects.

We next systematically examined whether a similar increased gene impact variability pattern could be observed in real single-cell aging datasets. DKO was applied separately to each species, tissue and cell type, and the variability of each gene's predicted impact was calculated independently within young and aged populations. Each violin plot in **Fig. 2** represents the distribution of gene-level perturbation-response variability across the analyzed genes. We evaluated single-cell data across a diverse panel of cell types spanning four distinct species, including human ovary (23-29 vs. 49-54 years; **Fig. 2a, Fig. S2**), human skeletal muscle (15-20 vs. 70-75 years; **Fig. 2b, Fig. S3**), mouse tissues from Tabula Muris Senis (3 vs. 24 months; **Figs. 2c,d, Fig. S4**), *C. elegans* (day 3 vs. day 11; **Fig. 2e**) and Drosophila brain (day 0 vs. day 30; **Fig. 2f**). Age groups followed the definitions of the original studies. Detailed dataset information, including data sources, cell-type annotations, young and old cell numbers, and exact *P* values for all comparisons, is provided in **Table 1** and Supplementary **Tables S1-S6**. Across 45 tissue–cell-type comparisons, gene impact variability was higher in aged cells in 41 comparisons, with 40 reaching statistical significance. Four comparisons showed lower variability in aged cells, three of which were significant, while two comparisons were not significant. Thus, although the direction was not uniform across all cell types, the predominant pattern was an age-associated increase in gene impact variability, indicating that the predicted effect of perturbing the same gene becomes more heterogeneous across individual cells with age.

To assess whether this age-associated increase depended on the age group used for model training, we further performed a cross-age evaluation using mouse naive B cells, granulocytes, and promonocytes from bone marrow (**Fig. S5**). DKO models were trained exclusively on either young or aged cells, and each trained model was subsequently applied to both age groups without retraining. Under both training strategies, aged cells consistently exhibited higher gene-level perturbation-response variability than young cells. These results indicate that the observed age-associated increase is robust to the choice of training age group, rather than arising from training separate models for young and aged populations.

These results demonstrate a consistent aging-associated increase in gene-level perturbation-response variability across the examined biological contexts, supporting the transcriptional deregulation hypothesis at the gene level.

**Somatic mutation burden is associated with increased cell impact variability**

The accumulation of somatic mutations is a common feature of aging[1,4,16]. A fundamental question is whether the increase in impact variability observed during aging is linked to mutational damage accumulation, or whether these represent independent characteristics of aging. Because somatic mutations occur stochastically in individual cells, their accumulation may alter cellular regulatory states and thereby change the predicted impacts of multiple gene perturbations. Here, we asked whether greater somatic mutation burden is associated with increased cell impact variability, reflecting greater divergence of each cell's gene-impact profile from a low-mutation reference state (see **Methods**). To test this association, within each dataset or donor, cells were ranked by mutation burden and divided into low-, mid-, and high-mutation groups. The low-mutation group was used to establish a gene-specific reference impact profile, defined by the median predicted impact of each gene. Cell impact variability was then calculated as the summed absolute difference between each cell's gene impacts and the corresponding reference values (see **Methods**), and directly compared between the mid- and high-mutation groups.

We first evaluated this relationship using the GRN dynamical simulations described above. Cells without regulatory damage ($p = 0$) were used to establish the gene-specific baseline impact profile, against which cell impact variability was calculated at increasing damage probabilities ($p = 0.05$, 0.10 and 0.15). Cell impact variability increased progressively with regulatory damage, both when calculated from ground-truth knockout impacts (**Fig. S1d**) and when estimated using DKO-predicted impacts (**Fig. S1e**).

We next examined human pancreatic glucagon cells from five participants (**Fig. 3a**). Within each participant, cells were ranked according to the fraction of mutations attributed to the study-defined S1 mutational signature, which accumulates with aging in pancreatic cells^12. Because the contrast between mid- and high-mutation cells depends on the within-participant spread of S1 mutation burden, we expected participants with broader S1 distributions to show larger differences in cell impact variability. The 50 cells with the lowest S1 mutation burden were used to establish the gene-specific reference impact profile, while the next 50 and the 50 cells with the highest burden were defined as the mid- and high-mutation groups, respectively. Cell impact variability was then calculated for the mid- and high-mutation cells relative to the low-mutation reference. In four of the five participants (P1-P3 and P5), high-mutation cells showed greater cell impact variability than mid-mutation cells. P4 showed the smallest difference between the two groups, consistent with the expected dependence of the effect on within-participant mutation-burden spread.

We further tested this association in cardiac fibroblasts from aged mice (**Fig. 3b, Fig. S6**). Because mutation burden was not provided as a predefined signature score in this dataset, we calculated a per-cell mutation burden by integrating the single-cell mutation-count and gene-expression matrices, using an expression-normalized mutation ratio to reduce transcription-dependent coverage bias (see **Methods**). Cells were ranked independently within each donor according to this mutation-burden score, and the same low-, mid-, and high-mutation grouping strategy was applied. The low-mutation cells were again used to establish the reference perturbation-response profile. Across donors, high-mutation cells generally showed greater cell impact variability than mid-mutation cells, although the magnitude of the difference varied among animals. Together, these results associate the accumulation of somatic mutations with increasing divergence of single-cell perturbation-response profiles from relatively low-damage cellular states.

These results suggest that the accumulation of regulatory alterations and somatic mutations is associated with increased cell impact variability, reflecting greater divergence of individual cells from the low-damage or low-mutation perturbation-response reference state.

### Cellular senescence is associated with increased gene and cell impact variability

Finally, we asked whether increased perturbation-response variability was associated with cellular senescence, a well-established aging-associated cellular state that accumulates with age and contributes to age-related tissue dysfunction[17,18]. We tested this association in two datasets: human umbilical vein endothelial cells[19] (HUVECs) and WI-38 cells[20] (**Fig. 4**). In the HUVEC dataset, we retained the four clusters defined in the original study, which represent a progression from a proliferative state in cluster 1, through intermediate states in clusters 2 and 3, to a senescent state in cluster 4 (**Fig. 4a**). We independently calculated senescence module scores for individual cells (see **Methods**) and observed significant differences among the four clusters, with cluster 1 showing the lowest senescence-associated transcriptional activity (**Fig. 4b**). Notably, the module score did not increase monotonically across the four clusters, with cluster 3 showing slightly higher values than cluster 4. This pattern is consistent with the multi-step nature of replicative senescence described in the original study, in which different senescence-associated transcriptional programs can peak at intermediate rather than terminal states. We first examined perturbation-response variability along this cluster progression. Gene impact variability increased from cluster 1 to 4, indicating progressively greater cell-to-cell heterogeneity in the predicted effect of perturbing the same gene (**Fig. 4c**). At the single-cell level, cluster 1 was used to establish the gene-specific reference impact profile. Cell impact variability subsequently increased from clusters 2 to 4, with the largest deviation observed in cluster 4 (**Fig. 4d**). Thus, although the senescence module score itself was not strictly monotonic, both gene- and cell-level perturbation-response variability increased along the proliferative-to-senescent progression defined in the original study.

To test this association using a direct measure of cellular senescence, we focused on cells within cluster 2 and grouped them by senescence module score. Cells were ranked by senescence module score, and the 50 cells with the lowest scores, cells ranked 51-100, and the 50 cells with the highest scores were selected as low-, mid-, and high-senescence groups, respectively (**Fig. 4e**). Gene impact variability was significantly higher in high-senescence cells than in low-senescence cells (**Fig. 4f**). Using the low-senescence cells to establish the reference impact profile, cell impact variability was also significantly higher in high-senescence cells than in mid-senescence cells (**Fig. 4g**). These within-cluster results further support an association between cellular senescence and increased heterogeneity and divergence of perturbation responses, independent of the original cluster assignments.

Then, we applied the same analysis to an independent WI-38 replicative-senescence dataset. Cells were sampled across six population doubling levels (PDLs 25, 29, 33, 37, 46 and 50) defined in the original study (**Fig. 4h**), and senescence module scores increased along the replicative-aging trajectory (**Fig. 4i**). Gene impact variability increased with PDL (**Fig. 4j**). Similarly, when PDL 25 cells were used to establish the reference impact profile, cell impact variability increased at later PDLs (**Fig. 4k**). We further focused on PDL 50 and selected low-, mid-, and high-senescence cells according to the same senescence-score ranking strategy (**Fig. 4l**). High-senescence cells showed significantly greater gene impact variability than low-senescence cells (**Fig. 4m**), while cell impact variability was significantly higher in high-senescence cells than in mid-senescence cells when the low-senescence group was used as the reference (**Fig. 4n**).

These findings indicate that cellular senescence is associated with increasing heterogeneity in gene-specific perturbation responses and greater divergence of single-cell perturbation-response profiles, consistent with a loss of regulatory robustness.

## Discussion

Transcriptional dysregulation has been proposed as a possible central mechanism contributing to functional decline during aging[21,22], but the evidence supporting its generality remains incomplete. Previous studies have provided evidence for this process through reduced gene co-expression in bulk transcriptomic data[23] and increased transcriptional variability at the single-cell level[11,24]. However, bulk analyses do not capture stochastic differences among individual cells, whereas early single-cell studies were often limited to specific cell types or tissues. More recent work has shown reduced global gene-to-gene transcriptional coordination across diverse organisms and cell types[14], providing broader evidence that transcriptional dysregulation is a general feature of aging. However, such global measures provide limited resolution regarding which genes contribute to this loss of coordination and how strongly individual cells deviate from a relatively stable regulatory state. Here, we extend this concept to a higher-resolution, perturbation-based perspective. Using DKO, we examined how predicted gene-perturbation effects vary across cells and how individual gene-impact profiles diverge from low-aging, low-damage, or low-senescence reference states, allowing transcriptional dysregulation to be examined at both gene and single-cell resolution.

Through systematic analyses of aging datasets from multiple species, we first found that the predicted effect of perturbing the same gene became increasingly variable across cells with age, indicating a loss of consistency in gene-specific regulatory responses. We then asked whether this age-related increase in impact variability was associated with two major features of aging: the accumulation of somatic mutations and cellular senescence. Cells with higher mutation burden showed greater cell impact variability, linking accumulated molecular damage to increased divergence of cellular gene-impact profiles. We next asked whether this regulatory instability was also related to cellular senescence, which may represent a downstream consequence of disrupted regulation rather than an upstream source of damage. Higher senescence levels were associated with increased gene and cell impact variability, both across senescence-associated populations and within the same cellular population. Together, these findings support a model in which somatic mutational damage is linked to increasing regulatory divergence, while cellular senescence may emerge as one consequence of this loss of regulatory stability. Our results therefore provide further evidence that transcriptional dysregulation is a general feature of aging.

An important implication of our findings is their potential to support the development of aging interventions. The age-associated increase in impact variability suggests that transcriptional dysregulation may reflect a broad loss of regulatory consistency across genes and cells, rather than disruption of only a small number of specific aging-related processes. Its association with somatic mutation burden further supports a link between accumulated molecular damage and this loss of regulatory stability. If stochastic damage contributes to the progressive increase in impact variability, interventions aimed only at individual aging mechanisms may therefore be insufficient, and strategies that slow damage accumulation or preserve regulatory robustness may also be important. At the same time, gene impact variability could be used to identify sets of genes whose perturbation responses change most strongly with age and determine which biological pathways and functions are most affected. These aging-related pathways could then provide candidates for functional testing and targeted intervention.

Several limitations should be considered. First, although DKO has been validated for predicting gene perturbation impacts, the impacts analyzed in the aging datasets are model predictions rather than direct perturbation measurements from the same cells. Second, the observed relationships between impact variability, aging, somatic mutation burden and cellular senescence are associative and do not establish causal direction. Mutation accumulation may contribute to regulatory divergence, whereas senescence may represent a consequence or correlate of regulatory instability. Third, because DKO models were trained separately for different tissues and cell populations, absolute impact values should not be directly compared across independently trained datasets; our conclusions therefore rely mainly on within-population comparisons. Finally, mutation burden estimated from scRNA-seq is incomplete because mutation detection depends on gene expression and transcript coverage. Although expression-aware normalization and within-donor comparisons reduce this bias, genomic measurements would provide a more complete assessment of somatic mutational damage.

In summary, DKO provides a practical approach for studying aging-related changes in gene regulatory responses at single-cell resolution without requiring a complete GRN or matched perturbation data. This makes it well suited to the large number of aging scRNA-seq datasets already available across species, tissues and cell types. Applying DKO to these datasets could reveal how gene perturbation effects change with age, identify genes showing the strongest age-associated changes, and test whether these genes overlap with known aging biomarkers or represent previously unrecognized candidates. Comparisons across tissues and cell types could further distinguish aging-related regulatory changes that are broadly shared from those that are cell-type specific. In this way, DKO may extend existing aging atlases from describing changes in gene expression to mapping how gene regulatory responses change during aging.

# Methods

## DKO model.

To map the expressed gene assemblage to their corresponding expression profiles, the binary gene set $\boldsymbol{z} \in \{0,1\}^N$ is initially transformed into a normalized representation $\tilde{\boldsymbol{z}} = \boldsymbol{z}/\mathbf{1}^{\mathrm{T}}\boldsymbol{z} \in \Delta^N$, where $\mathbf{1} = (1, \cdots, 1)^{\mathbf{T}}$. Subsequently, the gene assemblage $\boldsymbol{z}$ is processed through a two-hidden-layer multilayer perceptron (MLP) to capture complex data patterns:

$$f_w(\boldsymbol{z}) = w_2\big(w_1(\boldsymbol{z})\big),$$

where $w_1 \in \mathbb{R}^{(N_g \times N_g}$, $w_2 \in \mathbb{R}^{N_g \times N_g}$ represent the parameter matrices, and $N_g$ represent the number of genes. Next, Hadamard product and normalization are performed to compute the expression probability vector $\boldsymbol{p}$:

$$\boldsymbol{p} = \boldsymbol{z} \odot f_w(\boldsymbol{z}) / \sum \boldsymbol{z} \odot f_w(\boldsymbol{z}).$$

This formulation strictly constrains non-expressed genes to zero while ensuring that the total expression level across all genes sums to unity.

The MLP is trained to optimize the parameters by minimizing the following loss function over dataset $\mathcal{D}$:

$$E(\mathcal{D}) = \frac{1}{|\mathcal{D}|} \sum_{((\boldsymbol{z},\boldsymbol{p}) \in \mathcal{D})} d\big(\boldsymbol{p}, \hat{\varphi}_w(\boldsymbol{z})\big).$$

where $d$ specifies a dissimilarity or distance measure (e.g., Bray-Curtis dissimilarity). During training, all models were optimized for 5,000 epochs using a batch size of 20 and a learning rate of 0.001.

## Simulation damaged-related single cell data.

We generated synthetic single-cell data using a general GRN model[14] based on Michaelis–Menten dynamics[25]. The gene expression of each simulated cell was chosen to be the steady state of a coupled ordinary differential equation (ODE), which represent the dynamics of a group of genes with regulatory interactions. The gene expression state was randomly initialized such that 80% of the genes in each cell were expressed, with their initial expression levels sampled from a uniform distribution $\mathbb{U}(0.1,1)$, while the remaining 20% were unexpressed, i.e., their expression values were set to 0 at initialization and remained 0 throughout the dynamics. The ODE below describes the dynamics of gene $i$ expression levels in the model:

$$\frac{dx_i}{dt} = -B_i x_i + \sum_j w_{i,j}^{\text{act}} \frac{x_j}{1 + x_j} + \sum_j w_{i,j}^{\text{rep}} \frac{1}{1 + x_j},$$

$B_i$ is the degradation rate of gene $i$ (we set the degradation to be 1 for all genes). The activation and repression terms describe nonlinear regulatory inputs from other genes through weighted interactions.

All cells in a cohort or cell type shared the same GRN topology and a base weighted regulatory matrix $W^*$, with nonzero entries ($w_{(i,j)}^* > 0$) sampled from uniform distribution $\mathbb{U}(0,2)$. For each cell $v$, its weighted regulatory matrix $W^v$ was generated by perturbing $W^*$: each nonzero entry was retained with probability $1 - p$, or replaced by a new draw from uniform distribution $\mathbb{U}(0,2)$ with probability $p$. This procedure introduces cell-to-cell heterogeneity. The $p$ controls the level of cell-to-cell regulatory disruption: large $p$ perturbs a greater fraction of regulatory weights, corresponding to stronger damage and an aging-like deterioration of gene regulation within the cohort.

### Gene impact variability.

For each gene i, we quantified the variability of its perturbation impact across single cells using the median absolute deviation (MAD). Let $X_i = (x_{i,1}, \dots, x_{i,M})$ denote the vector of impact scores of gene i across M cells, where $x_{i,c}$ is the impact of gene i in cell c. The gene-wise median impact are

$$\widetilde{x_{\mathrm{i}}} = \text{median } (X_i).$$

The gene impact variability is then defined as the median absolute deviation from the gene-wise median:

$$GIV_i = \text{median}\left(\left|x_{i,c} - \widetilde{x_{\mathrm{i}}}\right|\right), \text{c} = 1, \dots, \text{M}.$$

A larger $GIV_i$ indicates greater cell-to-cell heterogeneity in the perturbation response of gene i.

### Cell impact variability.

We defined cell impact variability to quantify the extent to which the perturbation-response profile of an individual cell departs from a relatively stable reference state. To provide a common basis for comparison, cells with relatively low levels of aging, damage, or senescence were used to construct a gene-specific reference impact profile.

For each gene $i$, the baseline impact $b_i$ was defined as the median impact across cells in the corresponding reference population. Let $G$ genes in total, the baseline gene impact vector is $b = (b_1, \dots, b_G)^{\mathrm{T}}$. For one cell c, its gene-impact profile was represented as

$$y^{(c)} = \left(y_1^{(c)}, y_2^{(c)} \dots, y_G^{(c)}\right)^{\mathrm{T}},$$

where $y_i^{(c)}$ denotes the predicted impact of perturbing gene $i$ in cell $c$. The gene-wise deviation from the reference impact was defined as

$$z_i^{(c)} = y_i^{(c)} - b_i, i = 1, \dots, G,$$

Cell impact variability was then quantified as the summed absolute deviation across all genes,

$$\mathrm{CIV}(c) = \sum_{i=1}^{G} \left|z_i^{(c)}\right| = \sum_{i=1}^{G} \left|y_i^{(c)} - b_i\right|.$$

A larger value indicates that the cell's overall impact profile differs more strongly from the reference state.

### Cell mutation burden.

To quantify cell-type-specific mutation accumulation at single-cell resolution, we defined a per-cell mutation burden score by integrating single-cell mutation count matrix ($M$) and gene expression matrix ($E$) for each cell type. For cell $i$ and gene $j$, we first established an expression-normalized mutation ratio defined as

$$\mathrm{Ratio}_{i,j} = \frac{\ln\left(1 + M_{i,j}\right)}{1 + \ln\left(1 + E_{i,j}\right)}.$$

This ratio formulation accounts for transcription-driven coverage bias—where highly expressed genes naturally yield more reads and apparent mutations—while ensuring numerical stability when a mutated gene is unexpressed ($E_{i,j} = 0$). Next, to prevent non-mutated genomic regions ($M_{i,j} = 0$) from artificially diluting the estimate, we defined the per-cell mutation burden as the average ratio strictly across genes harboring detectable mutations:

$$\mathrm{Burden}_i = \frac{1}{\left|\{j \mid M_{i,j} > 0\}\right|} \sum_{j:M_{i,j}>0} \mathrm{Ratio}_{i,j}.$$

### Cell quality control procedure.

To ensure data quality, cell quality control was performed jointly across all cells within each cell type before age-group comparisons. (1) First, for each cell, genes with log-transformed expression $x_{i,j} > 0.2$ were defined as expressed, and the number of expressed genes was calculated for each cell. Cells with expressed-gene counts outside two standard deviations from the mean of the entire cell-type population ($\mu \pm 2\sigma$) were removed to exclude cells with unusually low or high gene detection. (2) Second, we calculated the group average expression vector and measured the Spearman distance ($d = 1 - \rho$, where ρ is the Spearman correlation coefficient) between each individual cell and this mean profile. We removed cells with a distance exceeding two standard deviations above the average distance ($d > \mu_d + 2\sigma_d$, where $\mu_d$ and $\sigma_d$ are the mean and standard deviation of distances to the mean profile).

### Senescence module scores calculation.

Senescence module scores were calculated for individual cells using a predefined set of senescence-associated genes, including *CDKN1A*, *CDKN2A*, *IGFBP3, IGFBP4, IGFBP5, IGFBP7, TIMP1, TIMP2, MMP1, MMP2, CCL2, PAPPA*, and *CTSB*. Scores were calculated using the "*scanpy.tl.score_genes*" function in python Scanpy package (version1.11.5). This approach quantifies the relative activity of the senescence-associated transcriptional program by comparing the average expression of the senescence gene set in each cell with that of background genes with comparable overall expression levels. This expression-matched background correction reduces the influence of differences in baseline gene abundance and provides a cell-level measure of coordinated senescence-associated transcriptional activity. Higher scores therefore indicate a stronger senescence-associated transcriptional state.

**Data availability.** All datasets analyzed in this study are publicly available. The corresponding studies, repositories, accession numbers, and download links for all datasets are provided in Supplementary Table S1.

**Code availability.** The code used in this work is available at https://github.com/yongjian66/DKO_keystone_genes.

**Acknowledgements.** X.-W.W. acknowledges the funding support from National Institutes of Health (K25HL166208).

**Author Contributions.** O.L. conceived the project. O.L. and X.-W.W. designed the project. Y.-J.H. performed all the numerical calculations and real data analyses. All authors wrote the manuscript. All authors contributed to the interpretation of the results and revised the manuscript.

**Competing Interests Statements.** The authors declare no competing interests. Correspondence and requests for materials should be addressed to O.L. (levyorr@biu.ac.il) and X.-W.W. (spxuw@channing.harvard.edu).

**Table 1. Datasets used to examine aging-associated regulatory and perturbation-response heterogeneity.**

| Species | Biological context | Tissues | Data modality | Study | State comparison | Role in this study |
|---|---|---|---|---|---|---|
| Human | Organismal aging | Ovary | snRNA-seq | Jin et al.[26], | Young vs Old | Cross-species replication |
| Human | Organismal aging | skeletal muscle | snRNA-seq / scRNA-seq | Kedlian et al.[27], | Young vs Old | Cross-species replication |
| Mouse | Organismal aging | Various tissues | scRNA-seq | The Tabula Muris Consortium[28] | Young vs Old | Discovery / aging analysis |
| *C.elegans* | Organismal aging | Various tissues | scRNA-seq | Roux et al.[29], | Young vs Old | Cross-species replication |
| Drosophila | Organismal aging | Brain | scRNA-seq | Davie et al.[30], | Young vs Old | Cross-species replication |
| Human | DNA damage | Pancreas glucagon cells | scRNA-seq | Enge et al.[31], | Low damage vs high damage | Damage-associated validation |
| Mouse | Mutation damage | Heart | scRNA-seq | The Tabula Muris Consortium[28] | Low damage vs high damage | Damage-associated validation |
| Human | Cellular senescence | WI-38 cells | scRNA-seq | Chan et al.[32], | Proliferating vs Senescent | Senescence validation |
| Human | Cellular senescence | human umbilical vein endothelial cells | scRNA-seq | Ahn et al.[19], | Proliferating vs Senescent | Senescence validation |

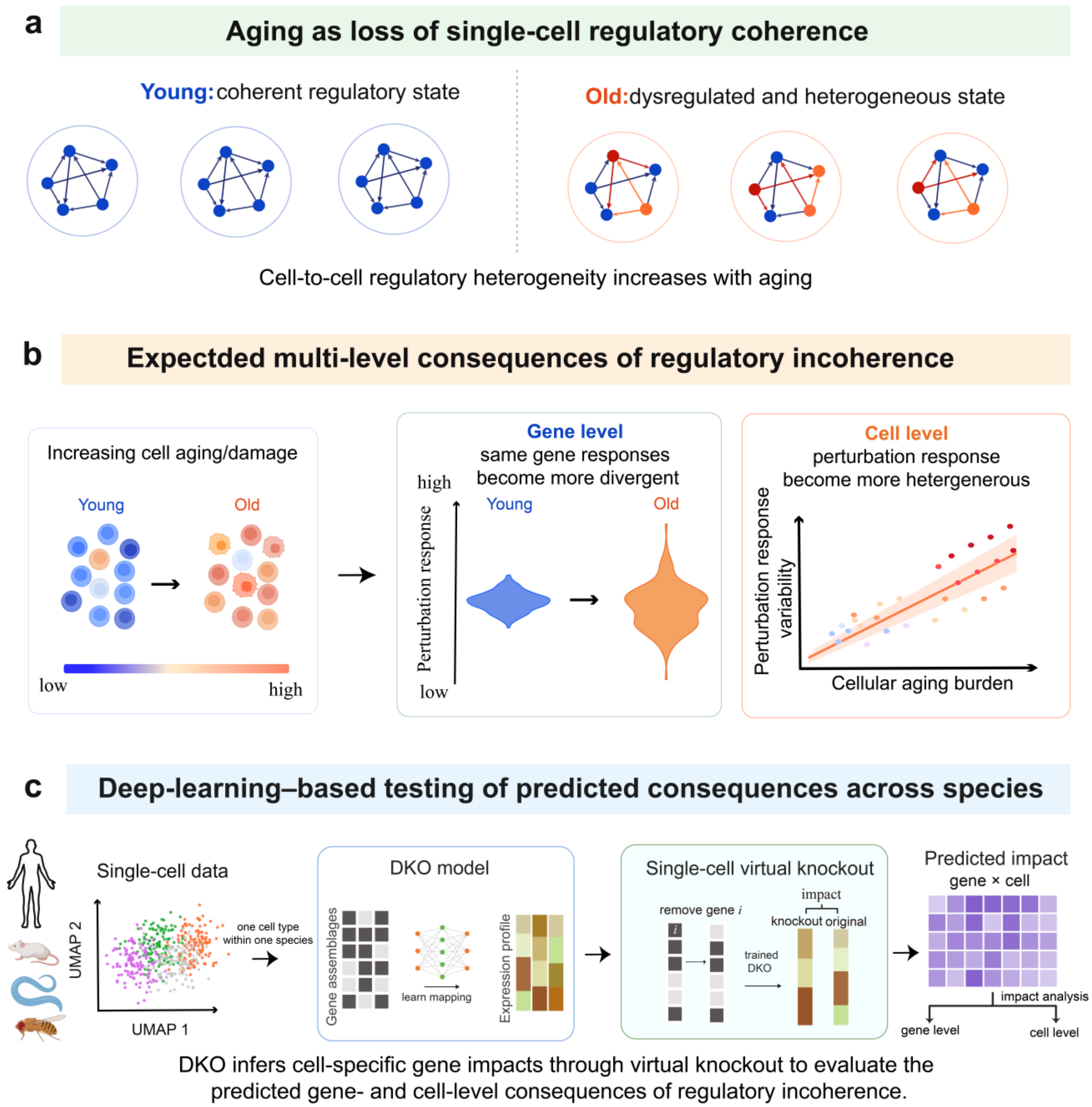

**Fig. 1. Conceptual and computational framework for linking aging-associated regulatory incoherence to heterogeneous perturbation responses. (a) Aging is conceptualized as a loss of single-cell regulatory coherence.** Young cells are expected to maintain relatively coherent gene regulatory states, whereas aging is associated with increasingly dysregulated and heterogeneous regulatory states across individual cells. **(b) Proposed multi-level consequences of regulatory incoherence.** Increasing cellular aging burden and stochastic molecular damage are expected to increase cell-to-cell divergence in responses to the same gene perturbation. At the gene level, the response to perturbing a given gene is predicted to become more heterogeneous across aged cells. At the cell level, cells with a higher aging burden are predicted to exhibit greater variability in their responses to gene perturbations. **(c) Deep-learning-based framework for testing these predictions across species.** Single-cell transcriptomic data from a defined cell type within each species are used to train a deep-learning DKO model that learns the mapping between

gene assemblages and cellular expression profiles. The trained model is then used to perform single-cell virtual gene knockouts, in which individual expressed genes are computationally removed from each cell and the resulting perturbation responses are predicted. Differences between the original and virtual-knockout expression profiles are quantified as gene-level impacts, generating a cell-by-gene perturbation impact matrix. This matrix enables quantification of perturbation-response heterogeneity at both the gene and cell levels and provides a framework for testing the predicted consequences of aging-associated regulatory incoherence.

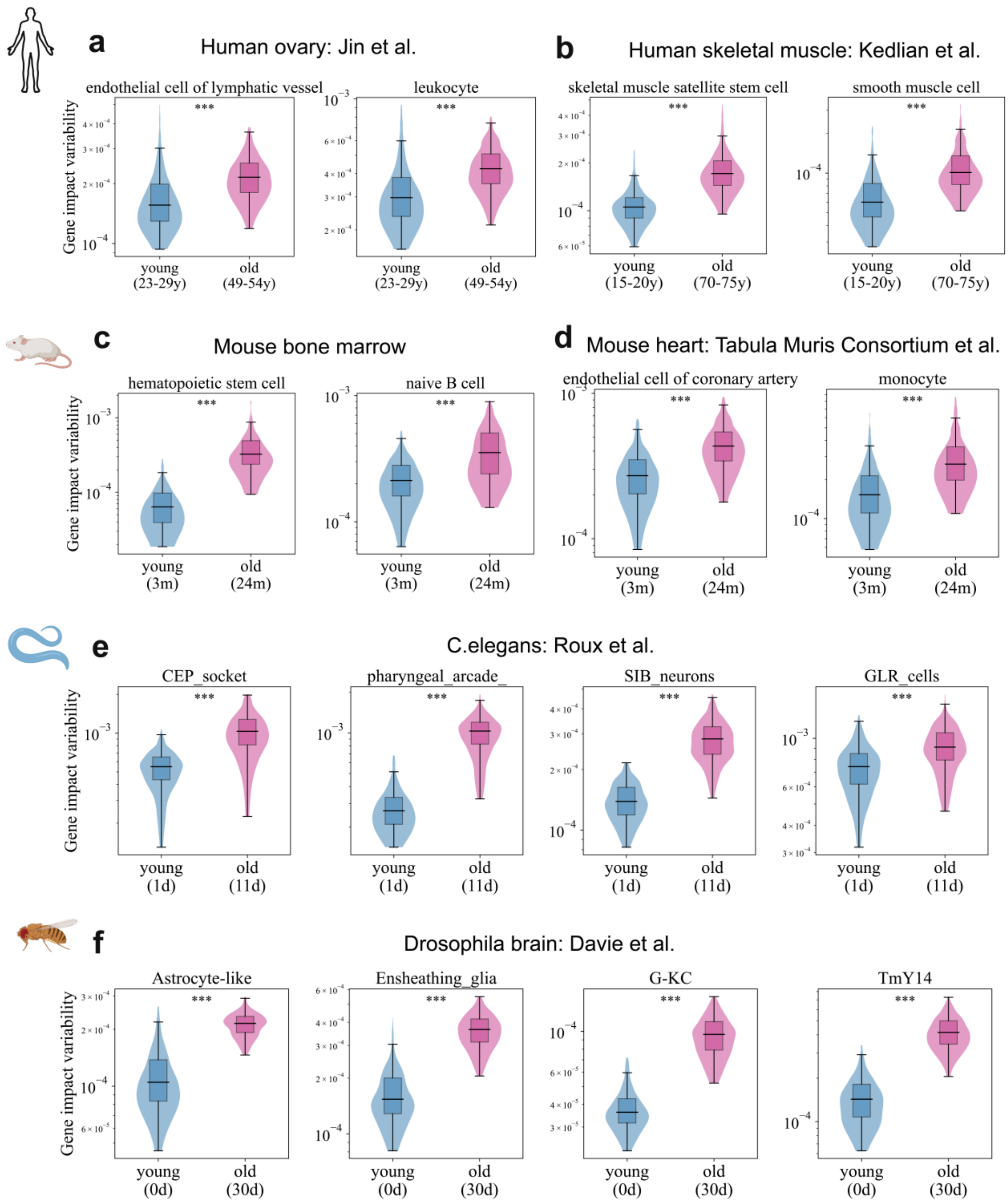

**Fig.2 Gene-level perturbation-response (defined as gene impact, see Methods) variability is increased in aged cells across multiple species, tissues, and cell types.** Each violin plot shows the distribution of gene impact variability between young and old cells within an indicated tissue-cell type population. Gene impact variability quantifies the heterogeneity of the predicted response to perturbation of the same gene across individual cells. **(a)** Human ovary, based on the single-nuclei multi-omics aging atlas reported by Jin et al.[26]; young and reproductively aged ovaries were

defined as 23-29 years and 49-54 years, respectively. (b) Human skeletal muscle, based on the aging atlas reported by Kedlian et al[27].; young and old individuals were defined as 15-20 years and 70-75 years, respectively. **(c,d)** Mouse bone marrow and heart from the Tabula Muris Senis dataset[33]; young and old mice were defined as 3 and 24 months, respectively. **(e)** *Caenorhabditis elegans*, based on the dataset reported by Roux et al[29].; young and aged animals were defined as 3 and 11 days, respectively. **(f)** *Drosophila* brain, based on the dataset reported by Davie et al[30].; young and aged flies were defined as 0 and 30 days, respectively. Analyses were performed separately for each tissue and cell type, using the corresponding age groups defined in the original datasets. Statistical significance was assessed using a two-sided paired Wilcoxon signed-rank test for gene-level comparisons between two groups. *P < 0.05, **P < 0.01, ***P < 0.001; ns., not significant.

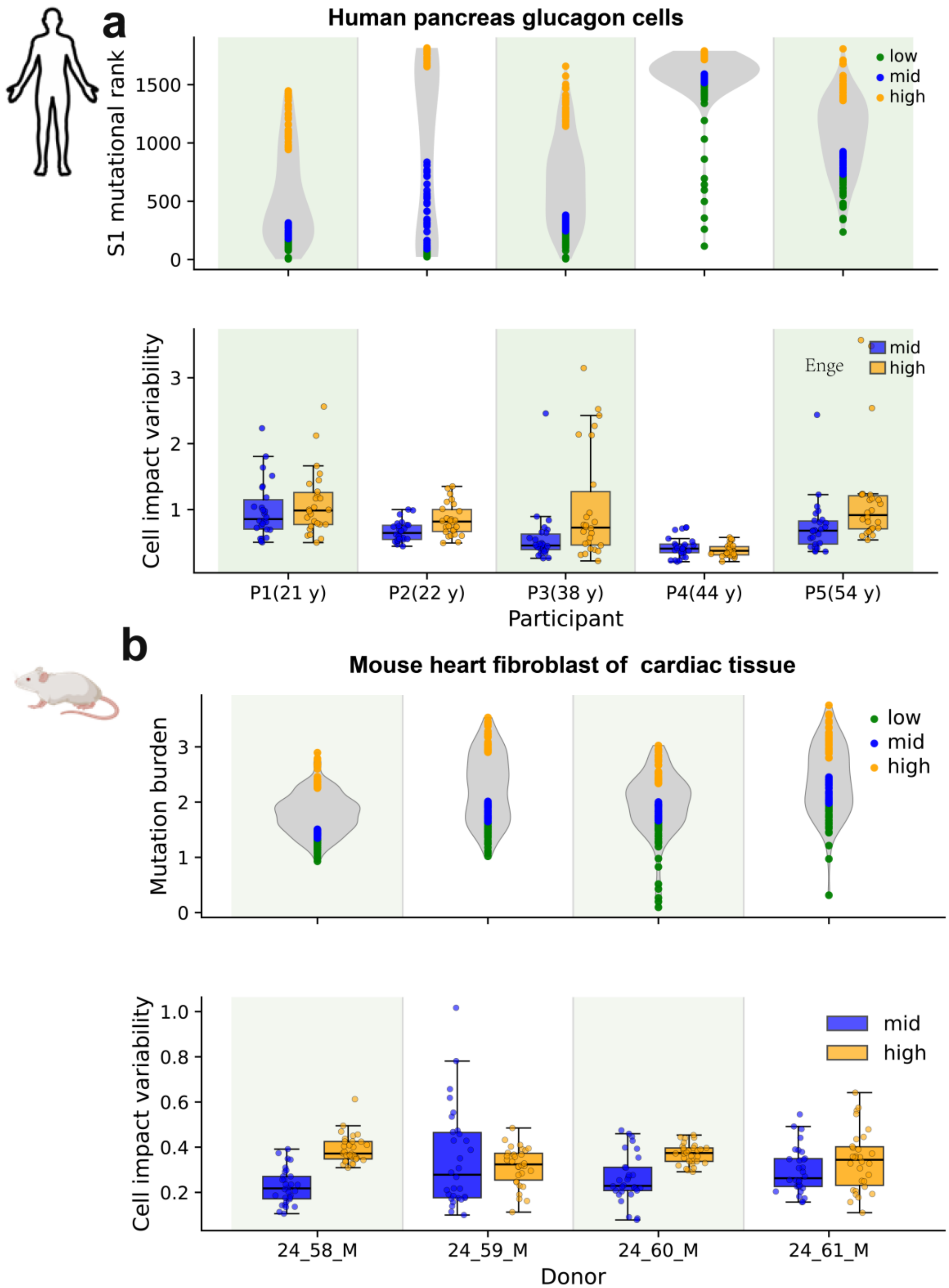

**Fig. 3. Somatic mutational burden is associated with increased cell-level regulatory impact variability in human pancreatic glucagon cells and mouse cardiac fibroblasts. (a)** Human pancreatic glucagon cells from five participants were ranked according to their burden of the S1 mutational signature (high rate of C > A, followed by C > G and C > T substitutions), as defined in ref. [31]. Cells were stratified into low-, intermediate-, and high-mutation groups according to their

within-participant mutational load distribution. Cell impact variability was then compared between cells with intermediate and high S1 mutational burden, using the low-mutation group as the reference baseline (see **Methods**). Across participants 1,2,3 and 5, cells with higher S1 mutational burden generally showed greater deviation from the low-mutation regulatory-impact baseline, indicating increased cell-level regulatory disruption with increasing somatic mutation burden. In contrast, the S1 mutational load in the cells of participant 4 is remarkably homogeneous; consequently, no significant deviation was observed. **(b)** Cardiac fibroblasts from aged mice were ranked by mutation burden separately within each donor and stratified into low-, intermediate-, and high-mutation groups. For each donor, the low-mutation cells were used to define a gene-specific baseline impact profile, calculated as the median predicted impact of each gene across low-mutation cells. For each intermediate- or high-mutation cell, cell impact variability was quantified as the summed absolute deviation of its gene-impact profile from this baseline (see **Methods**). High-mutation cells generally exhibited larger cell impact variability than intermediate-mutation cells across donors, supporting an association between increased somatic mutation burden and greater deviation of single-cell regulatory impact profiles from the low-mutation state.

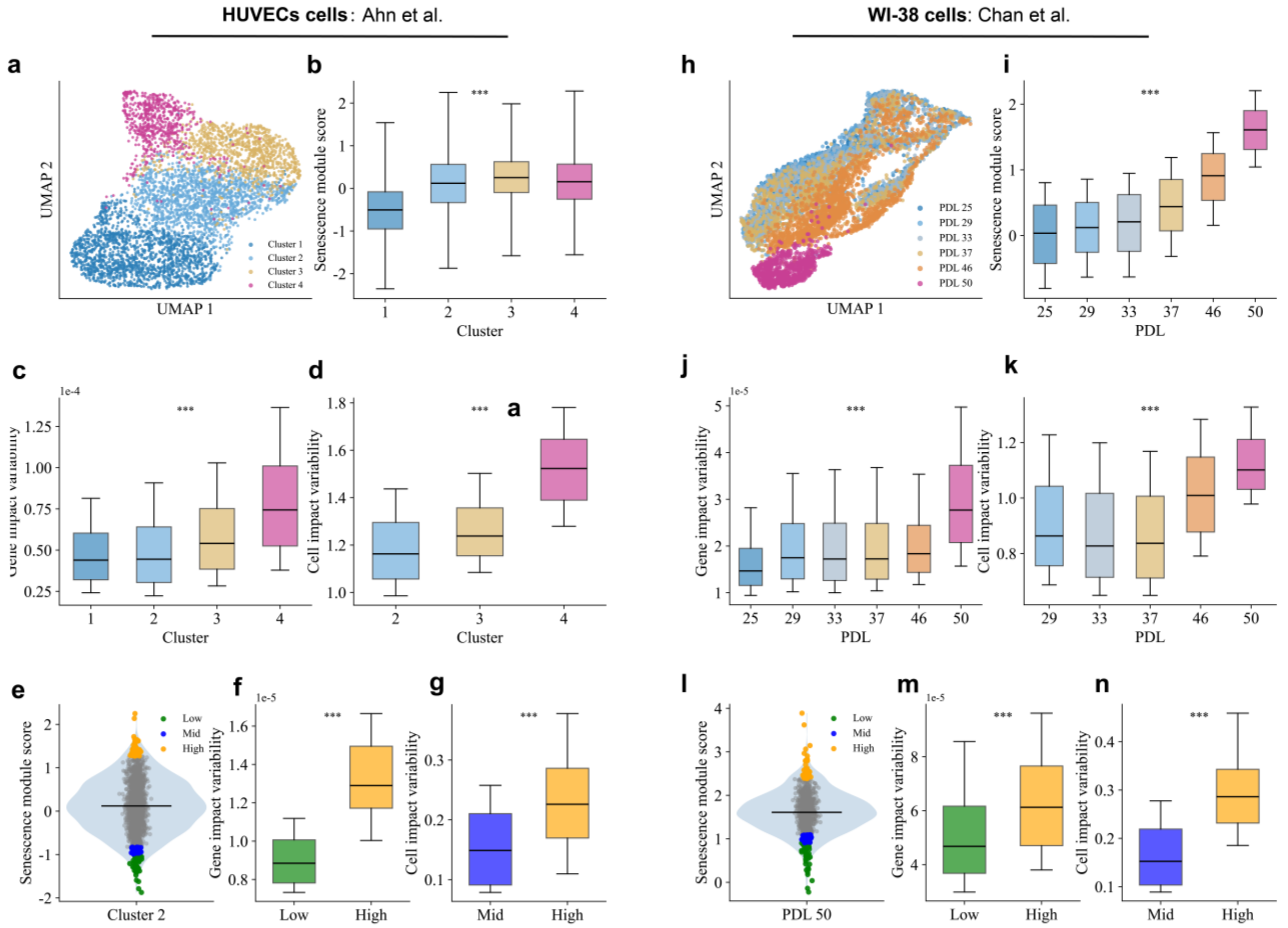

**Fig. 4. Senescence is associated with increased gene- and cell-level perturbation-response variability. (a-g)**, HUVEC cells. (**a**) UMAP visualization of the analyzed HUVEC cells, colored according to the four cluster assignments from the original study[19]. **(b)** Distribution of senescence module scores (see **Methods**) across HUVEC clusters. **(c)** Gene impact variability across the four clusters. **(d)** Cell impact variability in Clusters 2-4, calculated by comparing each cell's gene-impact profile with a gene-specific reference impact profile established from cluster 1 cells (see **Methods**). (**e**) Distribution of senescence module scores within cluster 2. After ranking cells by senescence score, the lowest 50 cells, cells ranked 51-100, and the highest 50 cells were highlighted as the low, mid and high groups, respectively. **(f)** Gene impact variability in low- and high-senescence cells within cluster 2. **(g)** Cell impact variability in mid- and high-senescence cells within cluster 2, calculated relative to a gene-specific reference impact profile established from low-senescence cells. **(h-n)** WI-38 cells. **(h)** UMAP visualization of WI-38 cells colored by population doubling level (PDL 25, 29, 33, 37, 46 and 50), with PDL annotations inherited from the original study[20]. **(i)** Distribution of senescence module scores across PDLs. **(j)** Gene impact variability across PDLs. **(k)** Cell impact variability at PDLs 29-50, calculated by comparing each cell's gene-impact profile with a gene-specific reference impact profile established from PDL 25 cells. **(l)** Distribution of senescence module scores within PDL 50. After ranking cells by senescence score, the lowest 50 cells, cells ranked 51-100, and the highest 50 cells were highlighted as the low, mid and high groups, respectively. **(m)** Gene impact variability in low- and high-senescence cells within PDL 50. **(n)** Cell impact variability in mid- and high-senescence cells

within PDL 50, calculated relative to a gene-specific reference impact profile established from low-senescence cells. Statistical significance was assessed using the Kruskal-Wallis test for comparisons among independent cell populations (**b, d, i, k**), the Friedman test for comparisons of the same genes across multiple groups (**c, j**), a two-sided paired Wilcoxon signed-rank test for gene-level comparisons between two groups (**f, m**), and a two-sided unpaired Mann-Whitney U test for cell-level comparisons between two groups (**g, n**). *P < 0.05, **P < 0.01, ***P < 0.001; ns., not significant.

# Supplementary Tables

**Table S1. Public data sources used in this study.**

| Species | Tissues | Study | Link to paper | Data Download link |
|---|---|---|---|---|
| Mouse | Heart, lung, … | The Tabula Muris Consortium[28] | https://www.nature.com/articles/s41586-020-2496-1 | https://cellxgene.cziscience.com/collections/0b9d8a04-bb9d-44da-aa27-705bb65b54eb |
| Human | skeletal muscle | Kedlian et al.[27], | https://www.nature.com/articles/s43587-024-00613-3 | https://cellxgene.cziscience.com/collections/2d40e6a7-f2fd-49ba-9db9-6b97e4c6dad5 |
| Human | Ovary | Jin et al.[26], | https://www.nature.com/articles/s43587-024-00762-5 | https://cellxgene.cziscience.com/collections/ce781ec4-393a-4902-94a0-5a1ca0b600ca |
| Drosophila | Brain | Davie et al.[30], | https://www.cell.com/cell/fulltext/S0092-8674(18)30720-7 | https://www.ncbi.nlm.nih.gov/geo/query/acc.cgi?acc=GSE107451 |
| *C.elegans* | Various tissues | Roux et al.[29], | https://www.cell.com/cell-reports/fulltext/S2211-1247(23)00913-0 | https://www.ncbi.nlm.nih.gov/geo/query/acc.cgi?acc=GSE208154 |
| Human | WI-38 cells | Chan et al.[32], | https://elifesciences.org/articles/70283 | https://www.ncbi.nlm.nih.gov/geo/query/acc.cgi?acc=GSE175533 |
| Human | HUVEGs | Ahn et al.[19], | https://www.sciencedirect.com/science/article/pii/S2405580825001293 | https://www.ncbi.nlm.nih.gov/geo/query/acc.cgi?acc=GSE98448 |
| Human | DNA damage | Enge et al.[31], | https://www.cell.com/cell/fulltext/S0092-8674(17)31053-X | https://www.ncbi.nlm.nih.gov/geo/query/acc.cgi?acc=GSE81547 |

**Table S2. Human ovary aging dataset used for gene impact variability analysis.** Human ovarian single-nucleus transcriptomic data were obtained from a single-nuclei multi-omics atlas of human ovarian aging reported by Jin et al.[26]. The original study profiled ovaries from four young individuals aged 23-29 years and four reproductively aged individuals aged 49-54 years. In the present study, we used the transcriptomic component of the dataset and retained the cell-type annotations provided by the original study without re-annotation. To ensure sufficient cell numbers for robust estimation, only cell types represented by more than 1,000 nuclei were retained for downstream analysis. Cell-level filtering (see **Methods** for a detailed description) was then applied separately within each selected population. The table reports the tissue, cell type, young and old age groups, the final number of cells retained in each age group after filtering, and the precise *P* values from a two-sided paired Wilcoxon signed-rank test comparing gene impact variability between young and old groups.

| Species | Tissue | Cell type | Ages details (young/old) | No. of cells (young/old) | Precise P values |
|---|---|---|---|---|---|
| Human | Ovary | endothelial cell of lymphatic vessel | 23-29/39-54 years | 476/193 | $2.797\times10^{-42}$ |
| Human | Ovary | leukocyte | 23-29/39-54 years | 536/409 | $1.230\times10^{-50}$ |
| Human | Ovary | Smooth muscle cell | 23-29/39-54 years | 657/604 | $5.171\times10^{-34}$ |

| Species | Tissue | Cell type | Ages details (young/old) | No. of cells (young/old) | Precise P values |
|---|---|---|---|---|---|
| Human | Ovary | Stromal cell | 23-29/39-54 years | 14,874/13,869 | $4.331\times10^{-1}$ |
| Human | Ovary | Theca cell | 23-29/39-54 years | 299/35 | $1.032\times10^{-173}$ |
| Human | Ovary | Blood vessel endothelial cell | 23-29/39-54 years | 712/594 | $2.383\times10^{-3}$ |

**Table S3. Human muscle aging dataset used for gene impact variability analysis.** Human skeletal muscle transcriptomic data this data generated by Kedlian et al[27]. Individuals aged 15-20 years were defined as the young group, whereas individuals aged 70-75 years were defined as the old group. Cell-type annotations provided by the original study were retained without re-annotation. Analyses were performed separately within each celltype. First, only cell types represented by more than 1,000 cells within a tissue were selected, resulting in 11 celltype populations. Cell-level filtering (see **Methods** for a detailed description) was then applied separately within each selected population. The table reports the tissue, cell type, young and old age groups, the final number of cells retained in each age group after filtering, and the precise *P* values from a two-sided paired Wilcoxon signed-rank test comparing gene impact variability between young and old groups.

| Species | Tissue | Cell type | Ages (young/old) | No. of cells (young/old) | |
|---|---|---|---|---|---|
| Human | Muscle | Fibroblast of the aortic adventitia | 15-20/70-75 years | 185/1371 | $5.290\times10^{-9}$ |
| Human | Muscle | Macrophage | 15-20/70-75 years | 527/1051 | $2.887\times10^{-257}$ |
| Human | Muscle | Capillary endothelial cell | 15-20/70-75 years | 84/1249 | $5.402\times10^{-72}$ |
| Human | Muscle | Muscle cell | 15-20/70-75 years | 184/1124 | $1.386\times10^{-145}$ |
| Human | Muscle | Vein endothelial cell | 15-20/70-75 years | 640/669 | $1.556\times10^{-34}$ |
| Human | Muscle | Pericyte | 15-20/70-75 years | 451/709 | $1.756\times10^{-256}$ |
| Human | Muscle | CD14-positive, CD16-low monocyte | 15-20/70-75 years | 179/650 | 1.559572e-$^{-66}$ |
| Human | Muscle | Endothelial cell of arteriole | 15-20/70-75 years | 904/502 | $3.216\times10^{-14}$ |
| Human | Muscle | Conventional dendritic cell | 15-20/70-75 years | 527/1051 | $1.635\times10^{-67}$ |
| Human | Muscle | skeletal muscle satellite stem cell | 15-20/70-75 years | 2706/4964 | $2.953\times10^{-165}$ |
| Human | Muscle | smooth muscle cell | 15-20/70-75 years | 2535/1252 | $2.120\times10^{-90}$ |

**Table S4. Mouse aging dataset used for gene impact variability analysis.** Mouse single-cell transcriptomic data were obtained from the Tabula Muris Senis dataset[28]. We restricted our analysis to the Smart-seq2-based dataset to minimize potential technical heterogeneity introduced by differences between single-cell sequencing platforms. Cell-type annotations provided by the original study were retained without re-annotation. Analyses were performed separately within each celltype. First, only cell types represented by more than 1,000 cells within a tissue were selected, resulting in 20 tissue-cell-type populations across 11 tissues. Cell-level filtering (see **Methods** for a detailed description) was then applied separately within each selected population. The table reports the tissue, cell type, young and old age groups, the final number of cells retained in each age group after filtering, and the precise *P* values from a two-sided paired Wilcoxon signed-rank test comparing gene impact variability between young and old groups.

| Species | Tissue | Cell type | Ages (young/old) | No. of cells (young/old) | Precise P values |
|---|---|---|---|---|---|
| Mouse | Bone marrow | hematopoietic stem cell | 3/24 months | 1249/919 | $9.213\times10^{-90}$ |
| Mouse | Bone marrow | naive B cell | 3/24 months | 620/945 | $6.010\times10^{-59}$ |
| Mouse | Bone marrow | Granulocyte | 3/24 months | 669/819 | $3.771\times10^{-17}$ |
| Mouse | Bone marrow | Promonocyte | 3/24 months | 316/321 | $6.093\times10^{-72}$ |
| Mouse | Heart | Fibroblast of cardiac tissue | 3/24 months | 1850/638 | $7.973\times10^{-2}$ |
| Mouse | Heart | endothelial cell of coronary artery | 3/24 months | 1038/900 | $2.599\times10^{-61}$ |
| Mouse | Heart | monocyte | 3/24 months | 411/353 | $2.851\times10^{-37}$ |
| Mouse | Large intestine | Epithelial cell of large intestine | 3/24 months | 797/207 | $3.256\times10^{-54}$ |
| Mouse | Large intestine | Large intestine goblet cell | 3/24 months | 566/409 | $7.973\times10^{-54}$ |
| Mouse | Limb muscle | Skeletal muscle satellite cell | 3/24 months | 472/43 | $5.236\times10^{-91}$ |
| Mouse | Lung | Bronchial smooth muscle cell | 3/24 months | 354/581 | $7.935\times10^{-2}$ |
| Mouse | Pancreas | Type B pancreatic cell | 3/24 months | 424/342 | $2.107\times10^{-3}$ |
| Mouse | Skin of body | Keratinocyte stem cell | 3/24 months | 1269/413 | $4.717\times10^{-107}$ |
| Mouse | Skin of body | Basal cell of epidermis | 3/24 months | 599/355 | $1.142\times10^{-32}$ |
| Mouse | Spleen | B cell | 3/24 months | 1063/708 | $4.004\times10^{-3}$ |
| Mouse | Thymus | DN4 thymocyte | 3/24 months | 455/500 | $2.115\times10^{-12}$ |
| Mouse | Thymus | Thymocyte | 3/24 months | 669/302 | $2.221\times10^{-4}$ |
| Mouse | Tongue | Basal cell of epidermis | 3/24 months | 1003/703 | $1.469\times10^{-117}$ |
| Mouse | Tongue | Keratinocyte | 3/24 months | 321/483 | $6.933\times10^{-109}$ |
| Mouse | Trachea | Fibroblast | 3/24 months | 517/308 | $7.424\times10^{-11}$ |

**Table S5. *C. elegans* aging dataset used for gene impact variability analysis.** Single-cell transcriptomic data from *C. elegans* were obtained from the original study[29]. We analyzed four cell types, with cell-type annotations retained from the original study. Cell-level filtering was applied separately within each cell type before downstream analysis. The table reports the analyzed cell types, young and old age groups, the final number of cells retained in each age group, and the exact *P* values from a two-sided paired Wilcoxon signed-rank test comparing gene impact variability between young and aged groups.

| Species | Cell type | Ages details (young/old) | No. of cells (young/old) | Precise P values |
|---|---|---|---|---|
| *C.elegans* | 5_0_CEP_socket | 1/11 days | 107/58 | $5.879\times10^{-96}$ |
| *C.elegans* | 15_0_pharyngeal_arcade_ | 1/11 days | 107/83 | $4.088\times10^{-191}$ |
| *C.elegans* | 22_0_SIB_neurons | 1/11 days | 126/72 | $1.180\times10^{-176}$ |
| *C.elegans* | 23_0_GLR_cells | 1/11 days | 142/97 | $2.191\times10^{-32}$ |

**Table S6. Drosophila aging dataset used for gene impact variability analysis.** Single-cell transcriptomic data from the Drosophila brain were obtained from the original study[30]. We analyzed four cell types, with cell-type annotations retained from the original study. Cell-level filtering (see **Methods** for a detailed description) was applied separately within each cell type before downstream analysis. The table reports the analyzed cell types, young and old age groups, the final number of cells retained in each age group, and the exact *P* values from a two-sided paired Wilcoxon signed-rank test comparing gene impact variability between young and aged groups.

| Species | Tissue | Cell type | Ages details (young/old) | No. of cells (young/old) | Precise P values |
|---|---|---|---|---|---|
| Drosophila | Brain | Astrocyte-like | 0/30 days | 83/168 | $1.452\times10^{-195}$ |
| Drosophila | Brain | Ensheathing_glia | 0/30 days | 163/154 | $8.930\times10^{-223}$ |
| Drosophila | Brain | G-KC | 0/30 days | 154/126 | $1.542\times10^{-228}$ |
| Drosophila | Brain | TmY14 | 0/30 days | 161/57 | $9.038\times10^{-230}$ |

# Supplementary Figures

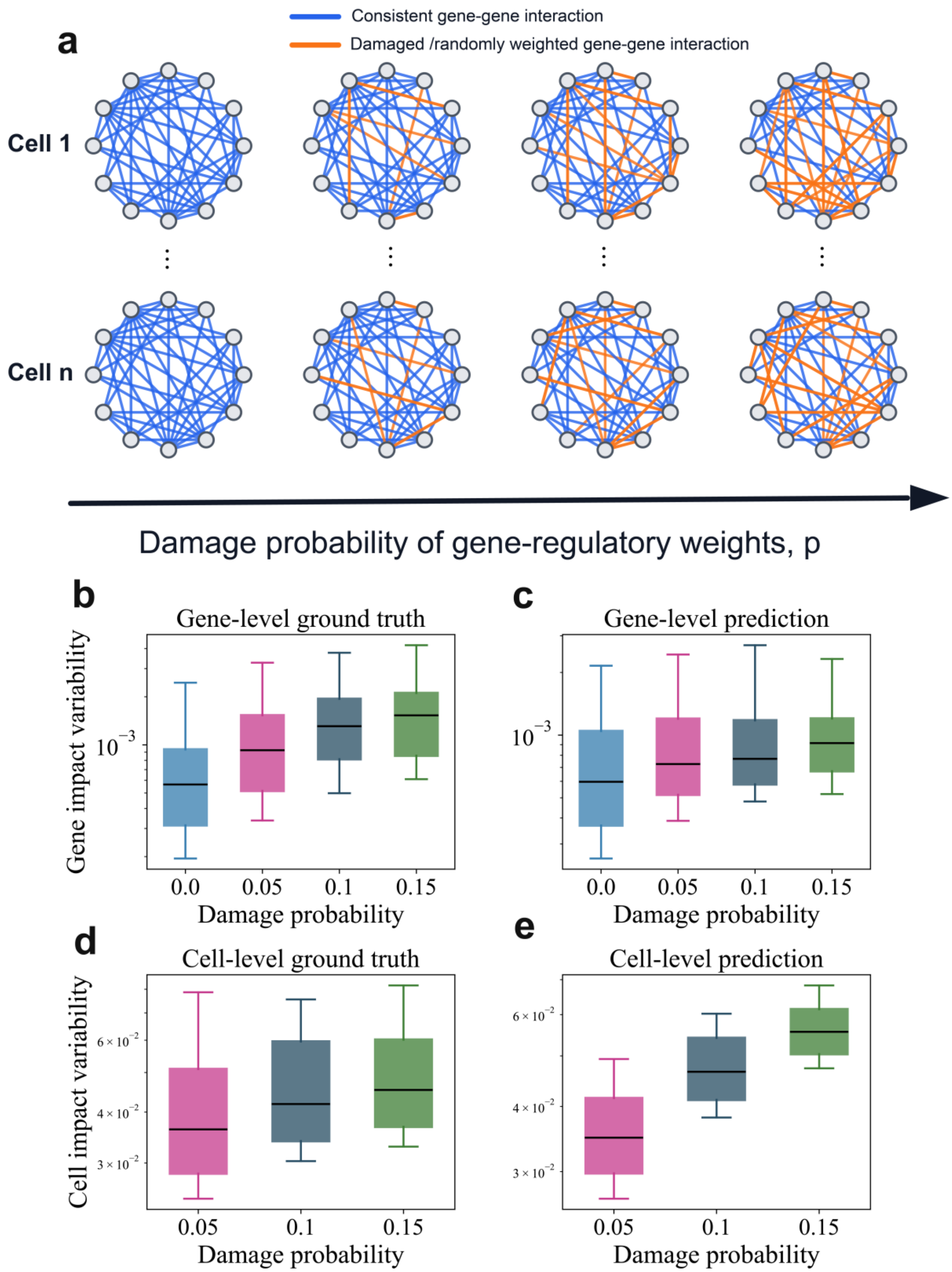

**Fig. S1. Simulated geneperturbation-response variability increases with aging.** (**a**) **Ground Truth.** In a dynamical simulation, with damage probability p ∈ 0,0.05,0.1,0.5, each cell's gene-regulatory weights are resampled at random (p = 0 means all cells share the same network). For each gene in each cell we perform an *in silico* knockout and define the cell-level impact as the Bray–Curtis dissimilarity between expression profile before and after knockout. For a given gene across cells, impact variability is quantified by the median absolute deviation (MAD). As p increases, cellular heterogeneity rises and the true impact variability increases. (b) **Prediction.** On the same simulated data, our DKO model predicts the increasing trend in impact variability. (c) **Real data.** Applying the DKO model to real data and comparing young vs old, we likewise observe higher gene mpact variability in the old group.

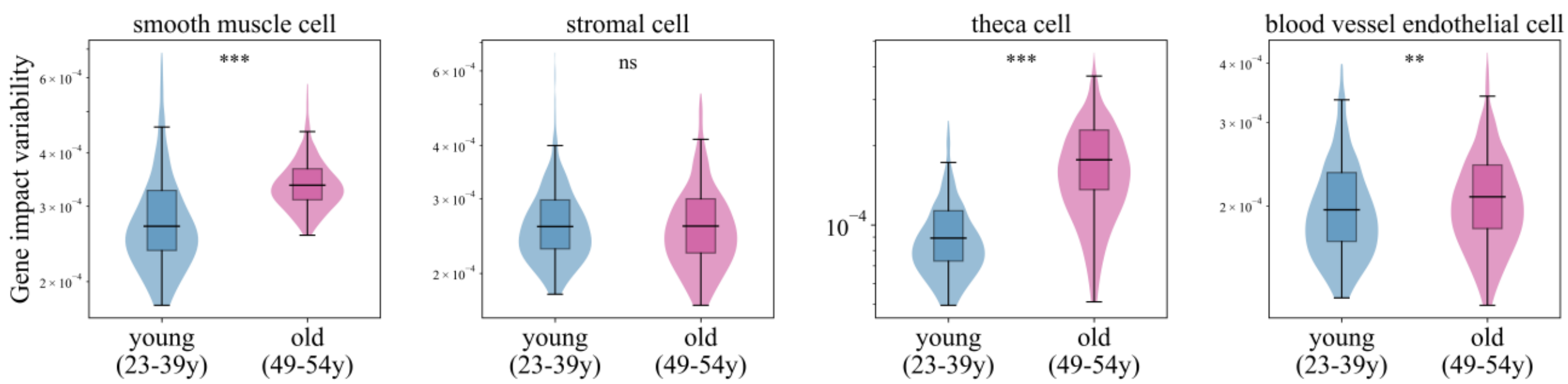

**Fig. S2. Gene impact variability in human ovary cell types.** Old cells (49-54 years) were compared to young cells (23-39 years) across four ovarian cell types. Statistical significance was assessed using a two-sided paired Wilcoxon signed-rank test for gene-level comparisons between two groups. *P < 0.05, **P < 0.01, ***P < 0.001; ns., not significant.

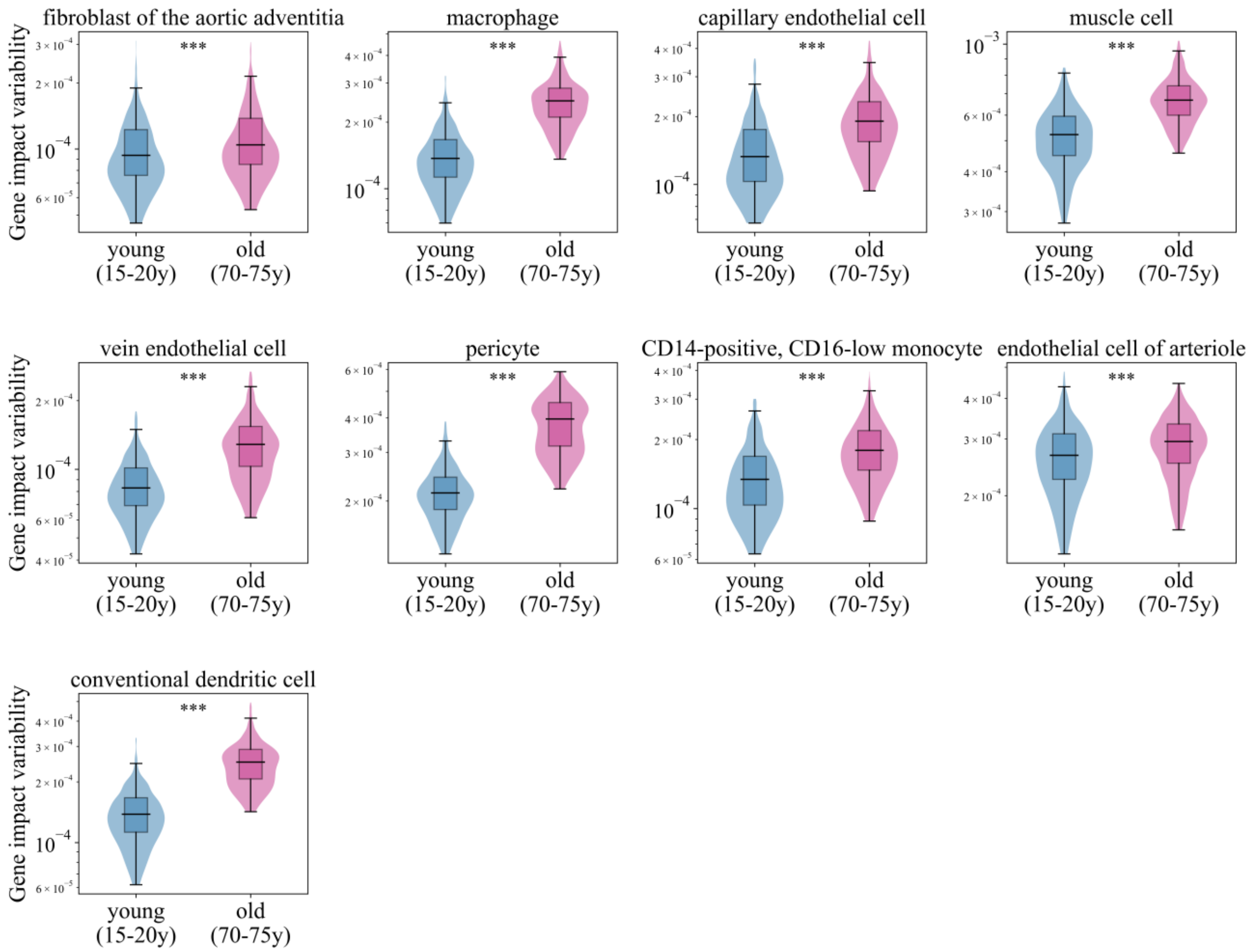

**Fig. S3. Gene impact variability in human muscle cell types.** Old cells (70-75 years) were compared to young cells (15-20 years) across nine muscle cell types. Statistical significance was assessed using a two-sided paired Wilcoxon signed-rank test for gene-level comparisons between two groups. *$P < 0.05$, **$P < 0.01$, ***$P < 0.001$; ns., not significant.

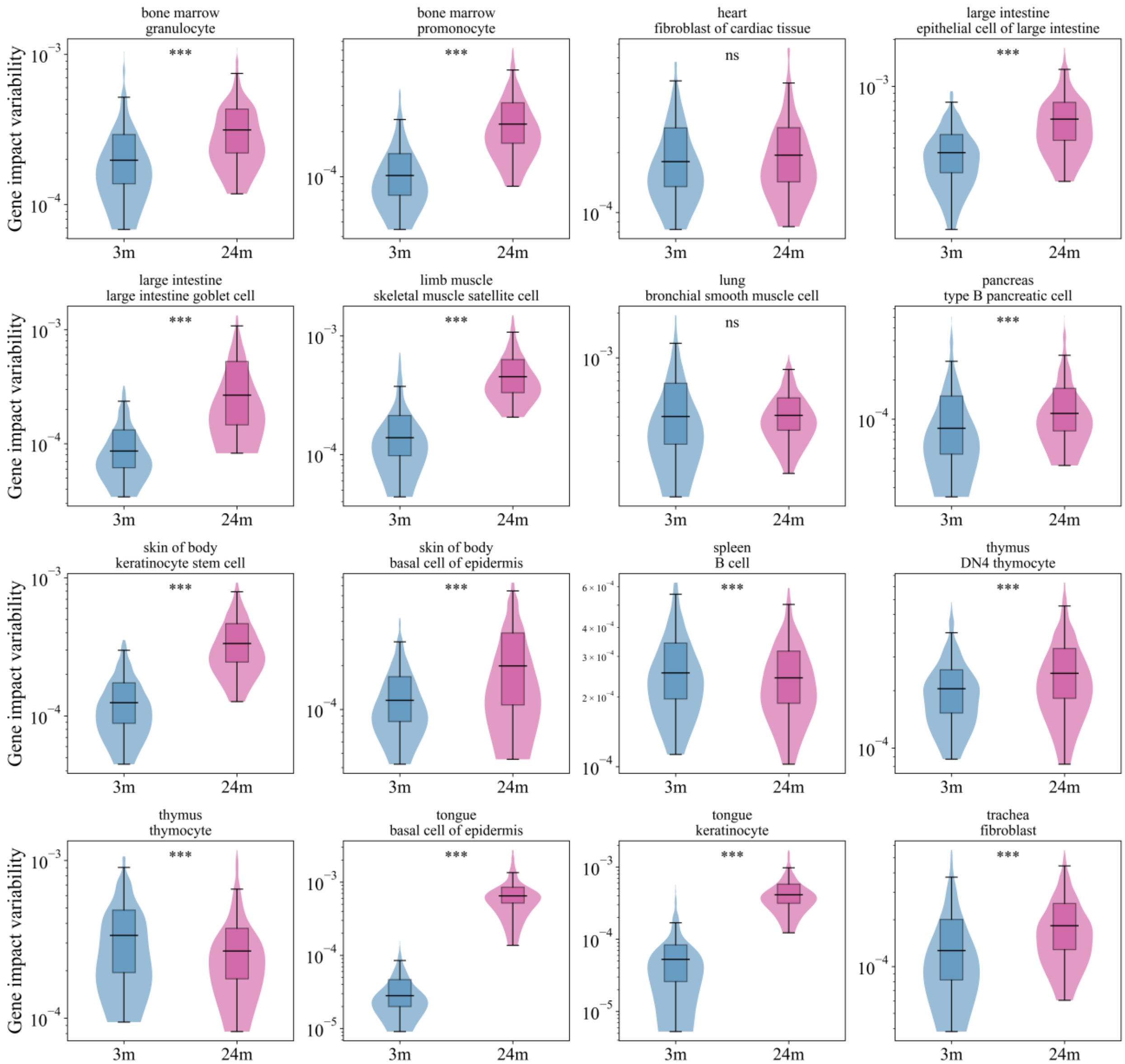

**Fig. S4. Gene impact variability in mouse cell types across tissues.** Old cells (24 months) were compared to young cells (3 months) across 16 cell types. Statistical significance was assessed using a two-sided paired Wilcoxon signed-rank test for gene-level comparisons between two groups. *P < 0.05, **P < 0.01, ***P < 0.001; ns., not significant.

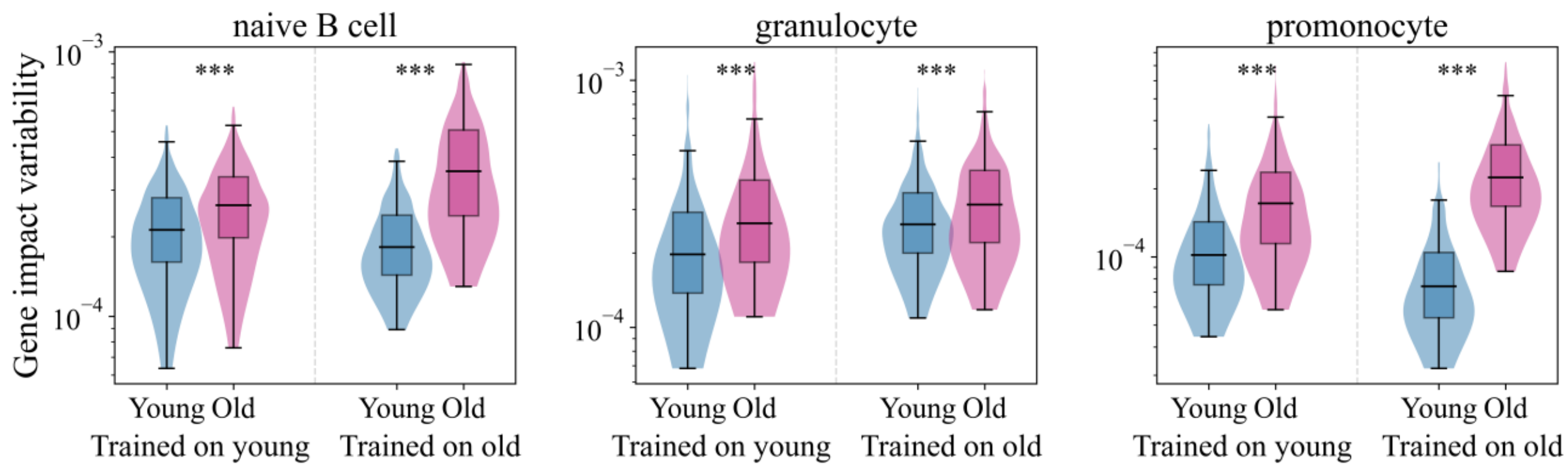

**Fig. S5. Robustness of the aging-associated increase in gene impact variability to training age.** To assess whether the aging-associated increase in gene impact variability was robust to the age group used for model training, DKO models were trained separately on young or old cells and then applied to both age groups without retraining. For each training strategy, gene impact variability was compared between young and old cells using the same trained model. Naive B cells, granulocytes, and promonocytes were analyzed as representative cell types. In the young-trained models, the model was trained exclusively on young cells and evaluated on both young and old cells. Conversely, in the old-trained models, the model was trained exclusively on old cells and evaluated on both young and old cells. In both training settings and across the examined cell types, old cells exhibited higher gene impact variability than young cells. These results indicate that the observed age-associated increase in perturbation-response heterogeneity is robust to the age group used for model training and is not dependent on training separate models for young and old cells.